\documentclass[preprintnumbers,amsmath,amssymb,12pt,floatfix,epsfig,A4]{revtex4-1}
\usepackage{bm}
\usepackage{natbib}
\usepackage[dvips]{color}
\usepackage{url}
\usepackage{graphicx}
\usepackage{amsmath}
\usepackage{float}
\usepackage{xcolor}
\date{\today}
\begin{document}

\title{Extended optical model analyses of $^{11}$Be+$^{197}$Au with dynamic polarization potentials}

\author{\textcolor{black}{Kyoungsu} Heo and Myung-Ki Cheoun}
\thanks{\textrm{e-mail:} cheoun@ssu.ac.kr}
\address{\textcolor{black}{Department of Physics and Origin of Matter and Evolution of Galaxy (OMEG) Institute, Soongsil University, Seoul 06978, Korea}}
\author{Ki-Seok Choi and K. S. Kim}
\address{School of Liberal Arts and Science, Korea Aerospace University, Koyang 10540, Korea}
\author{W. Y. So}
\address{Department of Radiological Science, Kangwon National University at Dogye,
Samcheok 25945, Korea}
\date{\today}

\begin{abstract}
 We discuss angular distributions of elastic, inelastic, and breakup cross sections \textcolor[rgb]{0.00,0.00,0.00}{for} \textcolor{black}{$^{11}$Be + $^{197}$Au system, which} were measured at energies below and around Coulomb barrier.
 To this end, we employ Coulomb dipole excitation (CDE) and long-range nuclear (LRN) potential to take into account long range effects by halo nuclear system  and break up effects by weakly-bound structure.
We then analyze recent experimental data including 3-channel\textcolor{black}{s} {\it i.e.} elastic, inelastic, and breakup cross sections, at $E_{\textrm{c.m.}}$=29.6 MeV and $E_{\text{c.m.}}$=37.1 MeV.
 From the extracted parameter sets using $\chi^{2}$ analysis,  we successfully reproduce the experimental angular distributions \textcolor{black}{of} the elastic, inelastic, and breakup cross sections for $^{11}$Be+$^{197}$Au system simultaneously. Also we discuss \textcolor{black}{the} necessity of LRN potential around Coulomb barrier from analyzed experimental data.
\end{abstract}

\pacs{24.10.-i, 25.70.Jj}
\maketitle

\section{Introduction}
\label{sec:org8292c5e}
Optical model (OM) is \textcolor{black}{a} traditional analysis tool for elastic scattering from \textcolor{black}{the} beginning of nuclear physics.
It provides valuable information of the nuclear interactions between projectile and target nucleus in \textcolor{black}{the} nuclear reaction.
Especially, \textcolor{black}{it is of} benefit to understand the elastic scattering and other direct reactions simultaneously by using $\chi^{2}$ analysis.
In practise, the parameters extracted from OM analysis are useful for other microscopic approaches such as distorted wave born approximation (DWBA) or continuum-discretized coupled-channels (CDCC) for treating contributions of inelastic and transfer channels.
Actually, elastic channel itself encompasses all participating direct reaction channel information, such as, inelastic, breakup, etc~\cite{satchler1983direct}.
Lots of efforts to separate each channel contribution from the elastic scatttering cross section have been continued. For example, it is well known that the dynamic polarization potential (DPP)\textcolor[rgb]{0.00,0.00,0.00}{~\cite{love1977dynamic}} related with E2 trans\textcolor[rgb]{0.00,0.00,0.00}{i}tion \textcolor{black}{is useful} for taking into account the contribution of inelastic channels in elastic scattering.
If we know all potentials corresponding \textcolor{black}{to} each nuclear reaction channel, we could simultaneously calculate cross sections of \textcolor[rgb]{0.00,0.00,0.00}{by each} nuclear reaction \textcolor[rgb]{0.00,0.00,0.00}{channel} with a proper parameter set.

We have been extending the OM \textcolor{black}{analysis by including} DPP which \textcolor{black}{comprises} potentials \textcolor{black}{from} breakup \textcolor[rgb]{0.00,0.00,0.00}{and fusion} reaction \textcolor{black}{channel}~\cite{PhysRevC.75.024610,PhysRevC.77.024609}.
Especially, the nuclear reactions including halo nuclei have been analyzed with the OM approach, \textcolor{black}{because those halo nuclei} are exhibiting obviously different behavior \textcolor{black}{from} known stable nuclei~\cite{PhysRevC.90.054615,PhysRevC.92.014627,PhysRevC.89.057601,PhysRevC.92.044618,PhysRevC.93.054624}.
For example, well-known light halo nuclei \textcolor{black}{such} as $^{11}$Li or $^{6}$He \textcolor{black}{have a specific feature owing to the halo nuclei composed by a} charged core and neutral one (two) valence neutron(s).
The charged core only \textcolor{black}{responses} to the Coulomb interaction by a target nucleus, but the valence neutron(s) \textcolor{black}{does not act} to the \textcolor[rgb]{0.00,0.00,0.00}{Coulomb one}.
\textcolor{black}{Consequently}, inelastic and breakup channel are opened \textcolor{black}{on} this weakly-bound structure.
These \textcolor{black}{phenomena} by Coulomb interaction \textcolor{black}{are} called as Coulomb dipole excitation (CDE)~\cite{AndresGomez-CamachoNagarajan1994,Gomez-CamachoAndresNagarajan1994,Andres199782}. \textcolor[rgb]{0.00,0.00,0.00}{The Coulomb breakup channels are such examples from these phenomena}. Furthermore, they also show \textcolor{black}{a} quite large distance of between the core nucleus and valence neutron(s) \textcolor{black}{termed as a} halo structure by \textcolor{black}{the} weak binding energy~\cite{TANIHATA1988592}.

\textcolor{black}{Recently}, interesting experimental researches including light neutron rich nuclei projectile such as $^{6,8}$He~\cite{PhysRevC.63.061603,PhysRevC.82.044617} ,$^{11}$Li~\cite{PhysRevLett.110.142701,Cubero2012,PhysRevC.87.044603,PhysRevC.92.044608} and $^{11}$Be~\cite{PhysRevLett.105.022701,PhysRevC.85.054607} and so on, have been carried out. \textcolor{black}{More recently, angular distribution data of $^{11}$Be+$^{197}$Au system including 3-channels {\it i.e.} elastic, inelastic and breakup cross sections around Coulomb barrier are reported~\cite{PhysRevLett.118.152502}.}
\textcolor{black}{Therefore, in this paper,} we \textcolor{black}{focus on the} nuclear reaction involving \textcolor{black}{$^{11}$Be} which is \textcolor{black}{a} well known halo 
nucleus \textcolor{black}{constructed} with a core nucleus $^{10}$Be and one weakly bound valence neutron.

In our \textcolor{black}{preceding} paper~\cite{PhysRevC.92.044618}, we already have calculated inelastic scattering for $^{11}$Be+$^{197}$Au system at $E_{\text{c.m}}$=30.1 MeV \textcolor{black}{below} Coulomb barrier (\textcolor{black}{$V_{B} \approx 37.9$ MeV}) by taking into account the first excitation state, \textcolor{black}{whose results} have shown good agreement with experimental data.
However, it has only contained inelastic channels without breakup cross section, which are involved in recent experimental data \cite{PhysRevLett.118.152502}.
Therefore, in this paper, we revisit the system and calculate angular distributions of $^{11}$Be on $^{197}$Au, \textcolor{black}{which} were measured at energies below and around Coulomb barrier, \textcolor{black}{$E_{\text{c.m.}}$=29.6 and 37.1 MeV}, with elastic, inelastic, and breakup channels.
In addition, we try \textcolor{black}{simultaneous} description of those elastic, inelastic and breakup cross section data with OM \textcolor{black}{approach by} considering long range DPPs  \textcolor{black}{such as CDE and long range nuclear (LRN) potential}.

The paper is organized as follows.
In Sec. 2, we \textcolor{black}{introduce} the optical model potential and its formalism implemented in the present work. In Sec. 3, we discuss the simultaneous \textcolor{black}{treatment} of elastic, inelastic and breakup cross sections using long range DPPs, such as LRN and CDE potential, with numerical results. Specifically, we discuss the necessity of LRN potential \textcolor{black}{for proper understanding angular distribution data of the reaction relevant to halo nuclei.} We finally summarize and conclude \textcolor{black}{our discussions} in Sec. 4.

\section{FORMALISM}
\label{sec:org830503a}
\subsection{Optical model potential}
\label{sec:org6ff0294}

For simultaneous analysis of all \textcolor{black}{participating} nuclear \textcolor{black}{reactions} in the scattering, the OM Schr\"odinger equation is expressed as follows~\cite{PhysRevC.65.044616,Kakuee2003339,MAHAUX1986354,PhysRevC.69.064606}:
\begin{equation} \label{scodinger}
[E - T_{l} (r)] \chi^{(+)}_{l} (r) = U_{\textrm{OM}} (r)~\chi^{(+)}_{l} (r),
\end{equation}
where $T_{l} (r)$ is \textcolor{black}{a} kinetic energy operator and $\chi^{(+)}_{l} (r)$ is \textcolor{black}{a} distorted partial wave function, respectively.
The kinetic operator is a function of the angular momentum $l$,
\begin{equation} \label{kinetic}
T_{l} (r) = - \frac{\hbar^{2}}{2 \mu} (\frac{d^{2}}{dr^{2}} - \frac{l (l + 1)}{r^{2}}),
\end{equation}
where $\mu$ and $l$ are \textcolor{black}{reduced} mass and \textcolor{black}{angular} momentum, respectively.
In the present calculation, the OM potential $U_{\textrm{OM}} (r)$ \textcolor{black}{comprises} the Coulomb potential $U_{\textrm{C}} (r)$, the nuclear potential $U_{\textrm{N}} (r)$, and the CDE potential $U_{\textrm{CDE}} (r)$ as follows:
\begin{eqnarray} \label{potential}
U_{\textrm{OM}} (r) &=& U_{\textrm{C}} (r) - U_{\textrm{N}} (r) - U_{\textrm{CDE}} (r)\nonumber \\
           &=& U_{\textrm{C}} (r) - [V_{\textrm{N}} (r) + i W_{\textrm{N}} (r)] \nonumber \\
           & & -[V_{\textrm{CDE}} (r) + i W_{\textrm{CDE}} (r)].
\end{eqnarray}
All \textcolor{black}{participating} potentials except Coulomb potential, $U_{\textrm{C}} (r)$, are consisted of \textcolor{black}{a} real part $V (r)$ and \textcolor{black}{an} imaginary part $W (r)$.
In general, \textcolor{black}{the} real part of potential has influence cross \textcolor{black}{sections} of elastic scattering, \textcolor{black}{while the} imaginary part is in charge of the absorption by each channel in OM approach.
\textcolor{black}{Therefore}, we assign \textcolor{black}{a different imaginary type to each channel.}

\subsection{Nuclear potentials}
\label{sec:org9d8af26}
\textcolor{black}{For} halo structure of  $^{11}$Be, we consider the nuclear potential, $V_{\textrm{N}} (r)$, in Eq.~(\ref{potential}) as coming from \textcolor{black}{two different nuclear interactions, which are between core part of projectile and target nucleus, and between one valence neutron part of projectile and target nucleus,} in the scattering of $^{11}$Be + $^{197}$Au system.

\textcolor{black}{First, for} the interaction between the core nucleus, $^{10}$Be, and the target nucleus, $^{197}$Au\textcolor{black}{, which} interaction is termed as the short-range (bare) nuclear potential (SRN) in this calculation, we employ a complex Woods-Saxon potential in a volume-type form. \textcolor{black}{In order to extract the short-range bare potential parameters for $^{11}$Be + $^{197}$Au system, we have to use the experimental elastic scattering data of $^{10}$Be + $^{197}$Au system. Unfortunately, however, there are no proper elastic scattering data at and above Coulomb barrier energy.}
\textcolor{black}{Thus,} potential parameters for SRN are \textcolor{black}{deduced from} the \textcolor{black}{$\chi^{2}$} analysis \textcolor{black}{by using the experimental elastic scattering data} of $^{10}$Be + $^{208}$Pb system~\cite{PhysRevC.69.047601} \textcolor{black}{and these extracted parameters will be replaced as the bare potential parameters of $^{10}$Be + $^{197}$Au system in this work.} The parameter set deduced from $^{10}$Be + $^{197}$Au system  is listed in \textcolor{black}{Table~\ref{parameters_10be}.}

 \begin{table}[h]
\begin{ruledtabular}
\begin{tabular}{cccccc}
 $V^{\textrm{sh}}_{0}$ &  $W^{\textrm{sh}}_{0}$  &  $a^{\textrm{sh}}_{0}$ & $a^{\textrm{sh}}_{W}$  &  $r^{\textrm{sh}}_{0}$ & $r^{\textrm{sh}}_{W}$\\
                   (MeV)        &          (MeV)         &            (fm)          &          (fm)    &            (fm)          &          (fm)    \\ \hline
                       113        &          169         &            0.63          &          0.30   &            1.06          &          1.20       \\
\end{tabular}
\end{ruledtabular}
\caption{The optical model parameters for SRN potential \textcolor{black}{for} $^{10}$Be + $^{208}$Pb system \textcolor{black}{adopted from} Ref. ~\cite{PhysRevC.69.047601}.
Here, $r_{i}$ = $R_{i}/(A^{1/3}_{1} + A^{1/3}_{2})$ with $i = 0 \; \mbox{and} \; W$ where 0 is real part and W is imaginary part.
$A_{1}$ and $A_{2}$ are masses of projectile and target nuclei, respectively.}
\label{parameters_10be}
\end{table}

\textcolor{black}{The other} interaction is the interaction between one valence neutron of projectile and target nucleus. \textcolor{black}{Strictly speaking, this interaction must also be considered as a part of the short-range bare potential for $^{11}$Be + $^{197}$Au system in the OM analysis, if the projectile were stable, or within the three-body system.}
But, one valence neutron in \textcolor{black}{the} projectile $^{11}$Be nucleus \textcolor{black}{is} easily detached by Coulomb and nuclear interactions generated from target nucleus, and the breakup reaction occurs mainly around the verge of \textcolor[rgb]{0.00,0.00,0.00}{$^{11}$Be} nucleus during the reaction process. \textcolor{black}{Therefore, the interaction between one valence neutron of projectile and target nucleus may give rise to both features from the short-range (bare) potential and the long range DPP associated with the breakup effect occurring at a distance. In this work, however, we do not separate both the short-range bare potential and the long range DDP. Instead, we introduce a new} long-range nuclear potential (LRN) with a surface-type Woods-Saxon potential corresponding to the nuclear interaction between one valence neutron of projectile and target nucleus\textcolor{black}{, which is given as} follows:
\begin{equation} \label{surface-woods}
U^{\textrm{lo}}_{0}(r)= (V^{\textrm{lo}}_{0} + i W^{\textrm{lo}}_{0}) f(X^{\textrm{lo}}_{0}),
\end{equation}
where $f(X^{\textrm{lo}}_{0}) = 4a_{0}^{lo}\frac{df(X^{\textrm{lo}}_{0})}{dR^{lo}_{0}}$ with $X^{\textrm{lo}}_{0}$ = $(r-R^{\textrm{lo}}_{0})/a^{\textrm{lo}}_{0}$  and $R^{\textrm{lo}}_{0}$
= $r^{\textrm{lo}}_{0}$ (${A_{1}^{1/3}}$ + ${A_{2}^{1/3}}$).
\textcolor{black}{Here} the parameter set for LRN potential is determined by $\chi^{2}$ analysis of both data.

\textcolor{black}{Resultant } parameter sets for LRN potential in this calculation are tabulated in Table~\ref{parameters_11be_r}\textcolor{black}{, which are determined} by the variation of the radius, $r^{\textrm{lo}}_{0}$=$r^{\textrm{lo}}_{W}$, from 1.5 to 3.5 fm. \textcolor{black}{To find reasonable parameters,  we exploit a simultaneous fitting of the elastic and breakup cross section data and find the optimum values for the four adjustable parameters, that is, $V^{\textrm{lo}}_{0}$  , $W^{\textrm{lo}}_{0}$, $a^{\textrm{lo}}_{0} = a^{\textrm{lo}}_{W}$.}
\textcolor{black}{Note that} we assume that radius of real part and diffuseness, $r^{\textrm{lo}}_{0}$ and $a^{\textrm{lo}}_{0}$, are the same with  $r^{\textrm{lo}}_{W}$ and $a^{\textrm{lo}}_{W}$ \textcolor{black}{to reduce number of} the parameters. We found that the diffuseness parameter, $a^{\textrm{lo}}_{0}$ = $a^{\textrm{lo}}_{W}$, is decreased with the increase of the radius.

\begin{table}[H]
\begin{center}
\begin{ruledtabular}
\begin{tabular}{ccccccc}
   $E_{\textrm{c.m}}$(MeV)     &    set    & $V^{\textrm{lo}}_{0}$  &   $W^{\textrm{lo}}_{0}$&  $a^{\textrm{lo}}_{0}$ = $a^{\textrm{lo}}_{W}$ & $r^{\textrm{lo}}_{0}$=$r^{\textrm{lo}}_{W}$ & $\chi^{2}$ \\
     (MeV)     &           &    (MeV)   &    (MeV)    &       (fm)       &       (fm)       &           \\ \hline
     37.1      &    (A)    &  -0.712    &    0.0085    &      5.12        &       1.5       &   1.41   \\
               &    (B)    & -0.118    &    0.0016    &      5.03        &       2.5       &   2.41   \\
               &    (C)    & -0.036    &    0.0010    &      4.72        &       3.5       &   2.53   \\  \hline
     29.6    &    (A)      & -0.203    &    0.0094    &      7.58        &       1.5       &   0.73   \\
               &    (B)    & -0.078    &    0.0044    &      7.08        &       2.5       &   0.82   \\
               &    (C)    &  -0.028    &    0.0022    &      6.24        &       3.5       &   1.03   \\
\end{tabular}
\end{ruledtabular}
\caption{Parameter sets of LRN potential with a surface-type used in Eq.~(\ref{surface-woods}) for $^{11}$Be + $^{197}$Au system.
Here, $r^{\textrm{lo}}_{i}$ = $R^{\textrm{lo}}_{i}/(A^{1/3}_{1} + A^{1/3}_{2})$ with $i = 0 \; \mbox{and} \; W$.}
\label{parameters_11be_r}
\end{center}
\end{table}

Here, we shortly summarize the nuclear interaction potential which is divided into two parts, short-range and long-range interaction potential, as follows :
\begin{eqnarray} \label{N-potential}
U_{\textrm{N}} (r) &=& U^{\textrm{sh}}_{0} (r) + U^{\textrm{lo}}_{0} (r) \nonumber \\
           &=&  [V^{\textrm{sh}}_{0} (r) + i W^{\textrm{sh}}_{0} (r)] \nonumber \\
           &+&  [V^{\textrm{lo}}_{0} (r) + i W^{\textrm{lo}}_{0} (r)].
\end{eqnarray}
where subscript "sh" is SRN and "lo" is LRN potential.  Also V(r) is a real part and W(r) is an imaginary part, respectively.

\subsection{Coulomb dipole excitation potential}
\label{sec:orgb936997}
Owing to the weakly bound structure of $^{11}$Be, valence neutron in projectile is easily \textcolor{black}{detached from the} core $^{10}$Be, or projectile is excited due to Coulomb interaction between projectile and target nuclei.
These phenomena \textcolor{black}{showed up} as strong absorption at forward angle region in the angular distribution of elastic scattering, because Coulomb interaction is long range interaction.
Therefore, we have to consider another long range \textcolor{black}{potential by Coulomb interaction.}

To \textcolor{black}{take into account} the excitation and breakup effects in \textcolor{black}{the} long-range region by Coulomb interaction, we employ CDE potential, $U_{\textrm{CDE}}$, which is constructed \textcolor{black}{by} two parts
\begin{equation}
U_{\textrm{CDE}}=U^{\textrm{inel}}_{\textrm{CDE}}+U^{\textrm{br}}_{\textrm{CDE}}=(V^{\textrm{inel}}_{\textrm{CDE}}+iW^{\textrm{inel}}_{\textrm{CDE}})+(V^{\textrm{br}}_{\textrm{CDE}}+iW^{\textrm{br}}_{\textrm{CDE}}),
\end{equation}
where $U^{\textrm{inel}}_{\textrm{CDE}}$ \textcolor{black}{and} $U^{\textrm{br}}_{\textrm{CDE}}$ \textcolor{black}{are} CDE potential for \textcolor{black}{excitation and} breakup channel, respectively.

\textcolor{black}{Detailed forms of} $U^{\textrm{inel}}_{\textrm{CDE}}$ and $U^{\textrm{br}}_{\textrm{CDE}}$ \textcolor{black}{are summarized} as follows  \cite{AndresGomez-CamachoNagarajan1994,Andres199782,PhysRevC.90.054615,PhysRevC.92.014627,PhysRevC.89.057601,PhysRevC.92.044618,PhysRevC.93.054624} :
\begin{eqnarray} \label{potential_coulomb-inela}
U^{\textrm{inel}}_{\textrm{CDE}}(r) &=& \frac{4 \pi}{9} \frac{Z^{2}_{t} e^{2}}{\hbar v} \frac{B(E1;\varepsilon_{x}^{1st})}{(r-a_{0})^{2} r}
\nonumber \\
&\times& ~[g(\frac{r}{a_{0}}-1,\xi) + i f(\frac{r}{a_{0}}-1,\xi)]
\end{eqnarray}
and
\begin{eqnarray} \label{potential_coulomb-break}
U^{\textrm{br}}_{\textrm{CDE}}(r) &=& \frac{4 \pi}{9} \frac{Z^{2}_{t} e^{2}}{\hbar v} \frac{1}{(r-a_{0})^{2} r}
\int^{\infty}_{\varepsilon_{b}} d\varepsilon \frac{dB(E1)}{d\varepsilon}\nonumber \\
&\times& ~[g(\frac{r}{a_{0}}-1,\xi) + i f(\frac{r}{a_{0}}-1,\xi)]
\end{eqnarray}
with
\begin{equation}
f(\frac{r}{a_{0}}-1,\xi)=4\xi^{2}(\frac{r}{a_{0}}-1)^{2} \exp(-\pi \xi)K''_{2i\xi}[2\xi(\frac{r}{a_{0}}-1)] \nonumber\textcolor{black}{,}
\end{equation}
where $a_{0}$ is \textcolor{black}{a} distance of \textcolor{black}{the} closest approach in head-on collision and
$Z_{t}$ is \textcolor{black}{a} charge number of \textcolor{black}{target} nucleus. $K''$ is the second derivative of a modified Bessel function
and $\xi$ = $a_{0} \varepsilon$/$\hbar v$ is an adiabatic parameter~\cite{AndresGomez-CamachoNagarajan1994}.\\

The expression $g(\frac{r}{a_{0}}-1,\xi)$ in \textcolor{black}{Eqs}.~(\ref{potential_coulomb-inela}) and~(\ref{potential_coulomb-break}) is a real part of CDE potential
extracted from the dispersion relation~\cite{AndresGomez-CamachoNagarajan1994} given by
\begin{equation}
g(\frac{r}{a_{0}}-1,\xi)= \frac{P}{\pi} \int^{\infty}_{-\infty} \frac{f(\frac{r}{a_{0}}-1, \xi)}{\xi-\xi'}  d\xi' .
\end{equation}

\textcolor{black}{In Eq.~(\ref{potential_coulomb-break})}, we employ the Coulomb strength distribution $dB(E1)/d\varepsilon$ as a simple model proposed from Ref.\cite{NAKAMURA1994296} as follows:
\begin{eqnarray}
\frac{dB(E1)}{d\varepsilon}&=&S\frac{\exp(2\kappa r_{0})}{1+\kappa r_{0}}\frac{3\hbar^{2}}{\pi^2\mu}e^{2}(\frac{Z_{1}}{A_1})^{2}\frac{\sqrt{S_{n}}(\varepsilon-S_{n})^{3/2}}{\varepsilon^{4}}\nonumber\\
&=&N\frac{{\sqrt{S_{n}}(\varepsilon-S_{n})^{3/2}}}{\varepsilon^{4}}\textcolor{black}{,}
\end{eqnarray}
where $S$ is the spectroscopic factor having $1.0\pm0.2$ value and $\kappa = \sqrt{2\mu S_n}/\hbar$ is a wave number  related to neutron separation energy, $S_{n}$, respectively. \textcolor{black}{Reduced} mass is given as $\mu$, and $r_{0}$ is \textcolor{black}{a} radius of the nuclear potential with a square-well type related to the halo $^{10}\textrm{Be}+n$ system.

\textcolor{black}{One} can also find some dependencies \textcolor{black}{on} $S_{n}$ \textcolor{black}{as well as} a proportional normalization constant $N$ \textcolor{black}{introduced} Ref.~\cite{PhysRevC.92.014627}.
In this \textcolor{black}{work}, we fix \textcolor{black}{the} proportional normalization constant N \textcolor{black}{as} 3.1~\cite{PhysRevC.92.014627}.
\textcolor{black}{All} results \textcolor{black}{depend} on this normalization constant, but \textcolor{black}{it does change an overall scale, but no shape difference.}
For $S_{n}$, we employ experimental data, 0.501 MeV, in Ref.~\cite{NAKAMURA1994296}.

\subsection{Angular distribution}
\label{sec:org5e5b8b5}

To investigate \textcolor{black}{the} cross section \textcolor{black}{by each potential discussed above}, we \textcolor{black}{use} the following form ~\cite{PhysRevC.65.044616}:
\begin{equation}
\frac{d\sigma_{\textrm{i}}}{d\Omega}=\frac{ka_{0}}{16\pi}\frac{1}{\cos(\frac{\theta_{\textrm{c.m.}}}{2})\sin^{3}(\frac{\theta_{\textrm{c.m.}}}{2})}\sum_{l}~\frac{\pi}{k}(2l + 1)~T_{\textrm{i}; l},~~~i = \textrm{BU~and~inel}
\label{cross-section}
\end{equation}
with
\begin{equation}
T_{\textrm{BU}; l} = \frac{8}{\hbar v} \int^{\infty}_{0} |\chi^{+}_{l} (r)|^{2}~[W^{\textrm{br}}_{\textrm{CDE}} (r) + W^{\textrm{lo}} (r)]dr
\label{T-br}
\end{equation}
for breakup reaction, and
\begin{equation}
T_{\textrm{inel}; l} = \frac{8}{\hbar v} \int^{\infty}_{0} |\chi^{+}_{l} (r)|^{2}~[W^{\textrm{inel}}_{\textrm{CDE}} (r)]dr
\label{T-inel}
\end{equation}
for inelastic scattering.
\textcolor[rgb]{0.00,0.00,0.00}{Here,  $k$ is the wave number defined by $\sqrt{2\mu E_{\textrm{c.m}}}/\hbar$. The distance of the closest approach, $a_0$, in a head-on collision is defined by a classical relation}
\begin{equation}
b = \frac{l}{k} = \frac{a_{0}}{2}\cot(\frac{\theta_{c.m.}}{2}),
\end{equation}
where impact parameter $b$ is given \textcolor{black}{as a} function of $l$ and $\theta_{\textrm{c.m.}}$. In Eq.~(\ref{T-br}), \textcolor{black}{we} assume that long-range absorption part, $W^{\textrm{lo}}$, contributes to breakup process, \textcolor{black}{because only} one excite states is included in \textcolor{black}{the} inelastic process.

\section{Results}
\label{sec:org5a96329}
\subsection{Elastic and quasi-elastic scattering}
\label{sec:orgcd26b71}
\begin{figure}[h]
\begin{tabular}{cc}
\includegraphics[width=0.50\linewidth]{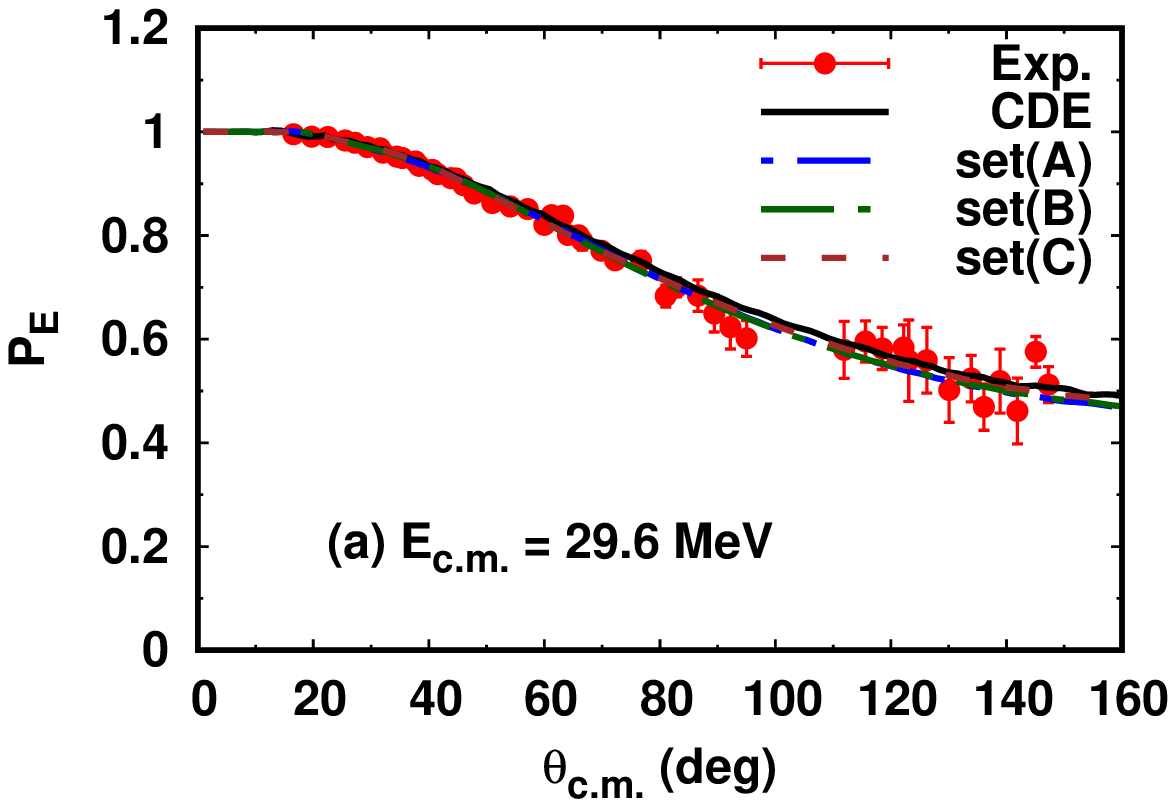} & \includegraphics[width=0.50\linewidth]{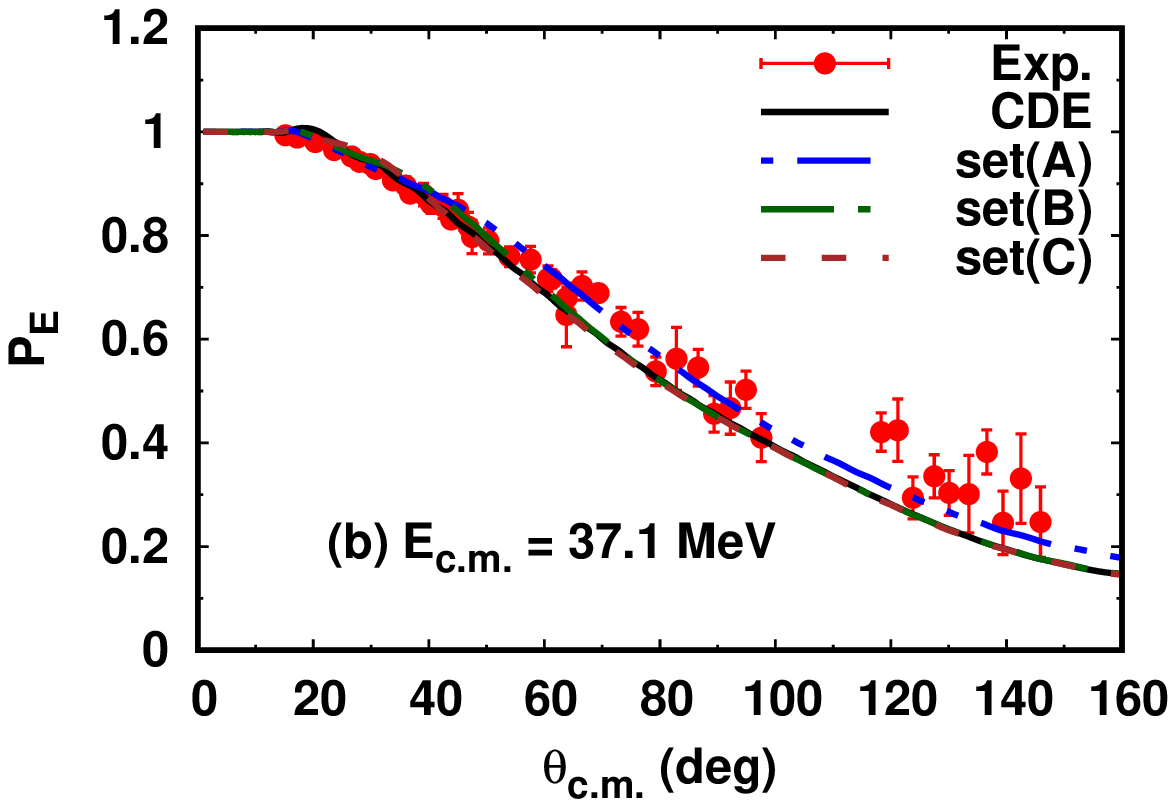} \tabularnewline
\end{tabular}
\caption{ (Color online)
\textcolor{black}{Differential} elastic scattering cross section ratio $P_{E}=\sigma_{\textrm{el}}/\sigma_{\textrm{RU}}$ of
$^{11}$Be + $^{197}$Au system at $E_{c.m.}$=29.6 MeV(\textcolor{black}{a}) and $E_{c.m.}$=37.1 MeV (\textcolor{black}{b}). The solid line is the ratio $P_{E}$
without the LRN potential and the dashed lines (blue, green and brown) are \textcolor{black}{obtained} by using the parameter sets in \textcolor{black}{Table}~\ref{parameters_11be_r}. Red circles are experimental elastic scattering data for $^{11}$Be + $^{197}$Au system taken from Ref.~\cite{PhysRevLett.118.152502}.}
\label{fig:1}
\end{figure}

First, we present results of elastic and quasi-elastic (QE) scattering cross \textcolor{black}{sections} using the long range dynamic polarization potentials (DPPs) \textcolor{black}{in} Eq. (\ref{potential}). To analyze elastic cross section data, we investigate the ratio of the elastic scattering cross section to Rutherford cross section, $P_{E}=\sigma_{\textrm{el}}/\sigma_{\textrm{RU}}$, in terms of a function of \textcolor{black}{center} of mass angle $\theta_{\textrm{c.m}}$.
\textcolor{black}{Figure} \ref{fig:1} presents the elastic scattering results given by the ratio $P_E$ at two \textcolor{black}{different} en ergies below and around Coulomb barrier, \textcolor{black}{$E_{\textrm{c.m}}$=29.6 and 37.1 MeV}, respectively.

\textcolor{black}{In Fig}.~\ref{fig:1}, the solid line means the ratio $P_{E}$ without the LRN potential, {\it i.e.}  $U_{\textrm{OM}}(r)=U^{\textrm{sh}}(r)+U_{\textrm{CDE}}(r)$, and the dashed lines (blue,green and brown) mean with the LRN potential, $U_{\textrm{OM}}(r)=U^{\textrm{sh}}(r)+U^{\textrm{lo}}(r)+U_{\textrm{CDE}}(r)$, using the parameters set (A), (B) and (C)  in \textcolor{black}{Table}~\ref{parameters_11be_r} respectively.

\textcolor{black}{In} Fig.~\ref{fig:1}, both results at $E_{\textrm{c.m}}=29.6$ MeV and $E_{\textrm{c.m}}=37.1$MeV show good and reasonable agreements with experimental data.
In the results at $E_{\textrm{c.m}}=29.6$ MeV (\textcolor{black}{Fig.~\ref{fig:1}(a)}), all sets including LRN potential show \textcolor{black}{almost} same results \textcolor{black}{albeit} different parameter sets.
We can \textcolor{black}{easily} notice that the LRN contribution \textcolor{black}{shows only a slight difference} with \textcolor{black}{the} solid line \textcolor[rgb]{0.00,0.00,0.00}{obtained without} LRN potential \textcolor[rgb]{0.00,0.00,0.00}{at} backward angles.
\textcolor{black}{Moreover, the effect of LRN potential at forward angle related to the long range region is shown to be insignificant compared to CDE potential result denoted as the solid black line.}
\textcolor{black}{Therefore}, we could not \textcolor{black}{conclude} whether the LRN potential is necessary or not \textcolor{black}{in these ratios for the elastic scattering} \textcolor{black}{in Fig.~\ref{fig:1}(a).} \textcolor{black}{In Fig.~\ref{fig:1}(b) for $E_{\textrm{c.m}}=37.1$ MeV, however, we noticed that the ratio $P_{E}$ obtained by adding LRN potential is significantly suppressed at forward angle (15$^{o}$ $\leq$ $\theta_{\textbf{c.m.}}$ $\leq$ 40$^{o}$) than that by CDE potential, although it is obscure in Fig.~\ref{fig:1}(b). It can be inferred that the strong breakup reaction by the LRN potential occurred at forward angle and the elastic scattering was reduced. Thus, this result demonstrates the necessity of LRN potential.}


\begin{figure}[h]
\begin{tabular}{cc}
\includegraphics[width=0.50\linewidth]{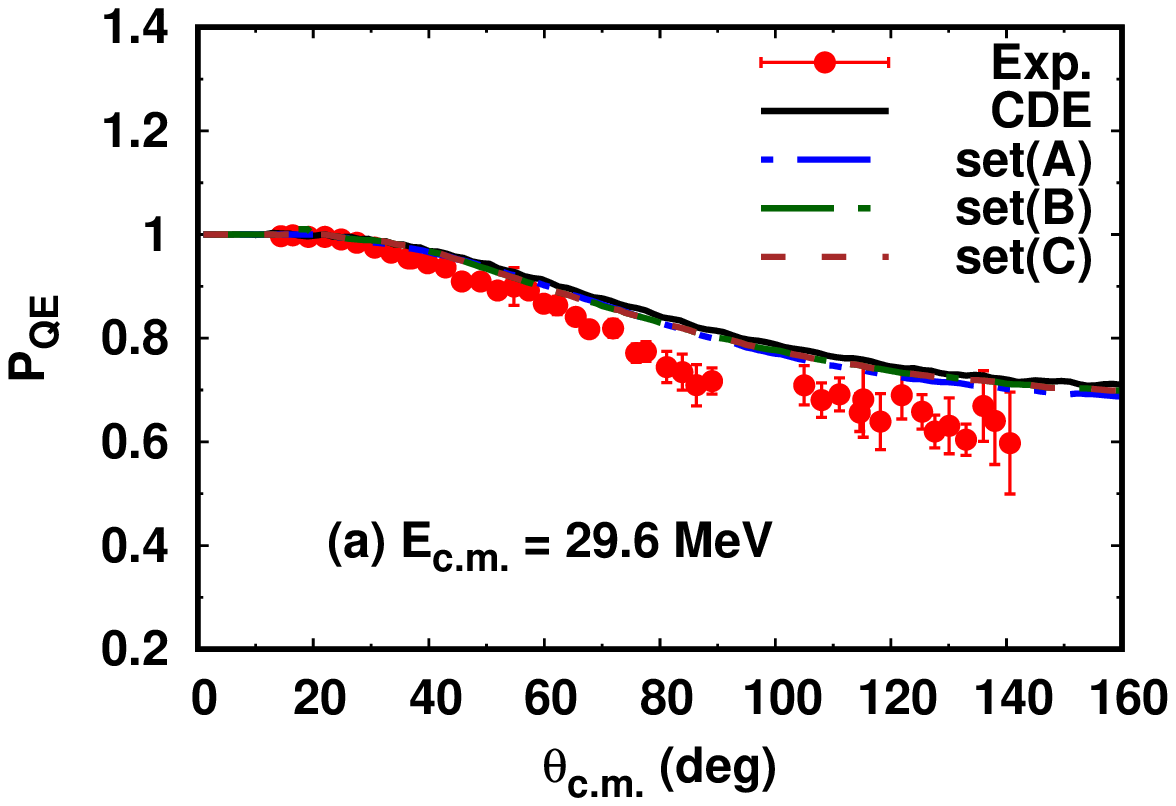} & \includegraphics[width=0.50\linewidth]{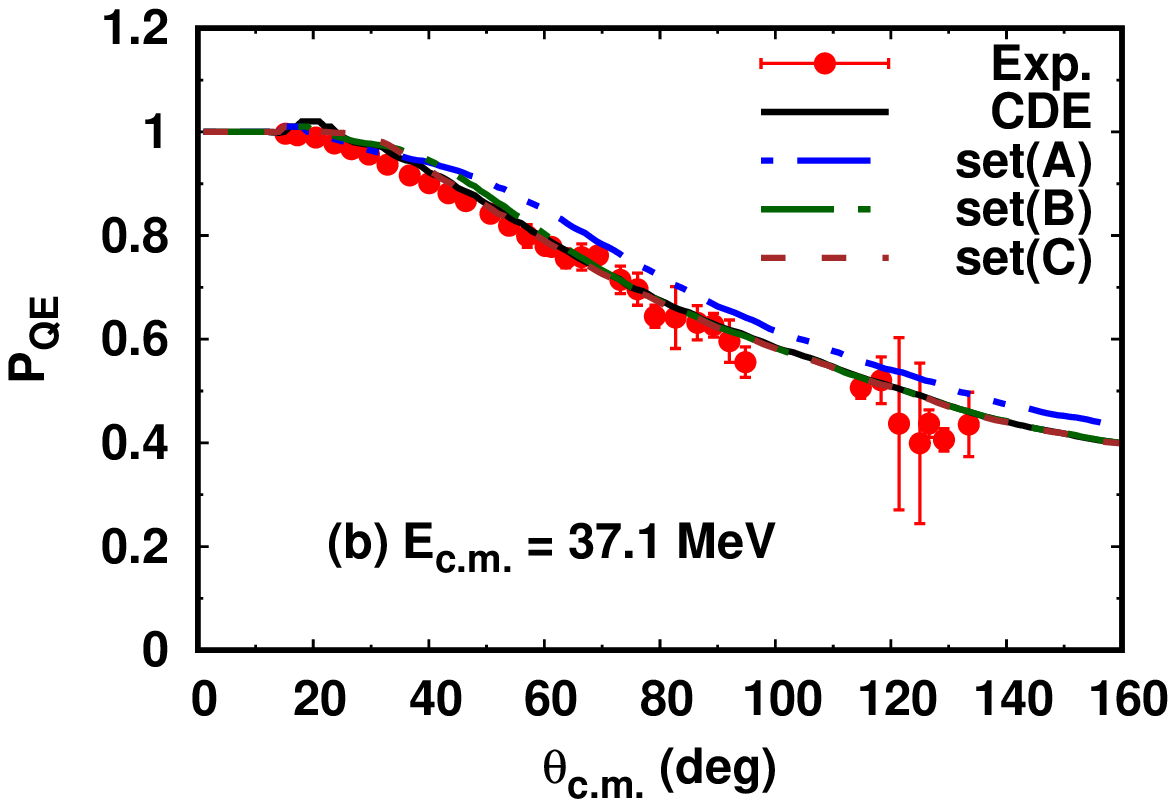} \tabularnewline
\end{tabular}
\caption{ (Color online)
The same as Fig.~\ref{fig:1} but for quasi-elastic (QE) scattering cross section. The experimental QE scattering data are taken from Ref.~\cite{borge2017scattering}.}
\label{fig:2}
\end{figure}

\textcolor{black}{Figure~\ref{fig:2}} presents QE scattering results with \textcolor{black}{the} same condition \textcolor{black}{as used in elastic scattering} in Fig.~\ref{fig:1}.
\textcolor{black}{Here,} QE scattering means \textcolor{black}{a} sum of elastic scattering and inelastic scattering cross section as defined in Ref.~\cite{borge2017scattering}.

For this calculation, we estimate inelastic 
\textcolor{black}{ cross section using the inelastic potential part in Eq.~(\ref{potential_coulomb-inela})} \textcolor{black}{and} add \textcolor{black}{it to the} \textcolor{black}{previous results from the elastic cross section shown in Fig.~\ref{fig:1}.} 
\textcolor{black}{In next subsection, the inelastic scattering results are compared with the experimental data.}
\textcolor{black}{From the viewpoint of QE channel, the remaining cross sections, excluding elastic and inelastic ones, will be the cross sections due to the breakup reaction and fusion channels.} 
\textcolor{black}{Since the} energy region \textcolor{black}{considered here} is \textcolor{black}{below and} around Coulomb barrier, \textcolor{black}{however,} \textcolor{black}{the contribution of the fusion cross section is not expected to be large at forward angle.}
It means that the absorption appeared in \textcolor{black}{the forward angle in Fig.~\ref{fig:2} almost stems from the} breakup contribution by Coulomb \textcolor{black}{and nuclear interactions.}

In the results of \textcolor[rgb]{0.00,0.00,0.00}{$E_{\textrm{c.m}}$} = 29.6 MeV(\textcolor{black}{Fig.~\ref{fig:2}(a)}), all \textcolor{black}{results slightly overestimated} experimental data. 
In the case of \textcolor[rgb]{0.00,0.00,0.00}{$E_{\textrm{c.m}}$} = 37.1 MeV(\textcolor{black}{Fig.~\ref{fig:2}(b)}), \textcolor{black}{however,} the whole results have good agreements \textcolor{black}{contrary to} the case of \textcolor[rgb]{0.00,0.00,0.00}{$E_{\textrm{c.m}}$} = 29.6 MeV. In \textcolor{black}{the} elastic scattering in Fig.~\ref{fig:1}, elastic cross \textcolor{black}{sections} at \textcolor[rgb]{0.00,0.00,0.00}{$E_{\textrm{c.m}}$} = 37.1 MeV have been underestimated, but those for quasi-elastic scattering \textcolor{black}{show much} better \textcolor{black}{results rather than those for elastic scattering}.


\subsection{Inelastic cross sections}
\label{sec:orgf44d747}
\begin{figure}[h]
\begin{tabular}{cc}
\includegraphics[width=0.50\linewidth]{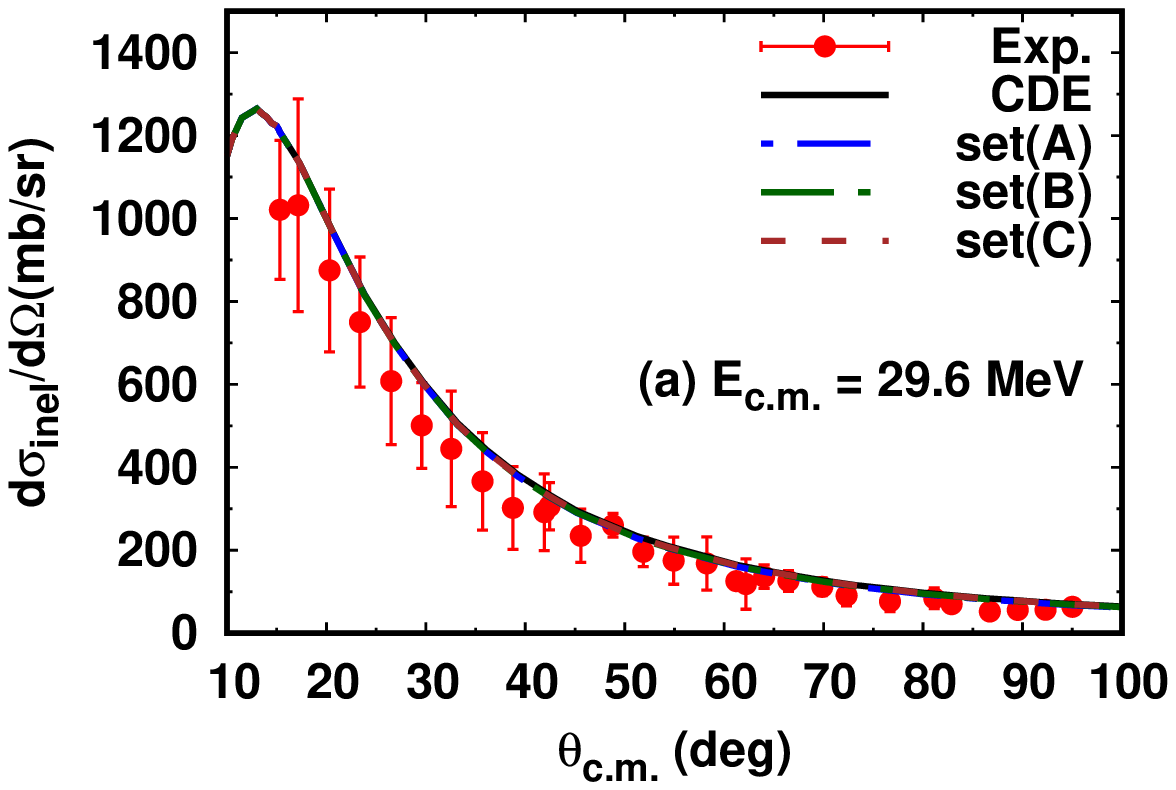} & \includegraphics[width=0.50\linewidth]{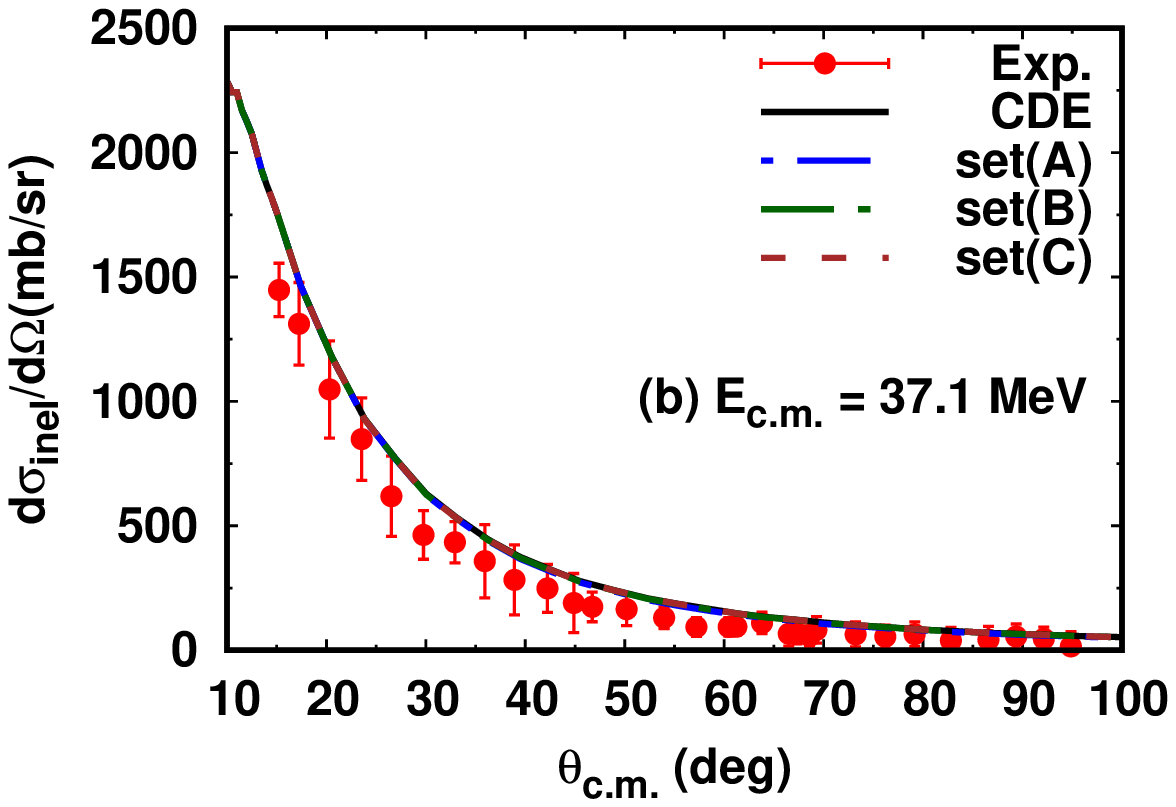} \tabularnewline
\end{tabular}
\caption{ (Color online)
\textcolor{black}{Inelastic} scattering cross sections \textcolor{black}{of} $^{11}$Be + $^{197}$Au system at $E_{\textrm{c.m.}}$=29.6 MeV(\textcolor{black}{a}) and $E_{\textrm{c.m.}}$=37.1 MeV (\textcolor{black}{b}). The \textcolor{black}{data are taken from} Ref.~\cite{PhysRevLett.118.152502}.}
\label{fig:3}
\end{figure}

In Fig.~\ref{fig:3}, we present \textcolor{black}{results} of angular distribution of inelastic scattering \textcolor{black}{using} the CDE potential introduced in Eq.~(\ref{potential_coulomb-inela}).
For the angular distribution, \textcolor[rgb]{0.00,0.00,0.00}{we use Eqs.~(\ref{cross-section}) and (\ref{T-inel})}.
\textcolor{black}{We take} the first excitation of $^{11}$Be\textcolor{black}{,} $\varepsilon^{1st}_{x} = 0.32$ MeV, with Coulomb dipole strength $B(E1;\varepsilon^{1st}_{x})$= 0.115 $e^{2}\textrm{fm}^{2}$\textcolor[rgb]{0.00,0.00,0.00}{~\cite{ajze98}}.
Because the neu\textcolor[rgb]{0.00,0.00,0.00}{t}ron separation energy of $^{11}$Be is 0.501 MeV, the projectile is broken up \textcolor{black}{for the} incident energy \textcolor{black}{considered here}.
The \textcolor{black}{target excitation in inelastic scattering} is also possible.
But, we ignore the target excitation because the \textcolor{black}{target} excitation \textcolor[rgb]{0.00,0.00,0.00}{states} \textcolor{black}{below one} neutron separation energy of $^{11}$Be \textcolor[rgb]{0.00,0.00,0.00}{are} almost E2 and E2+M1, \textcolor{black}{whose energy and contribution are} lower than E1.
Therefore, \textcolor[rgb]{0.00,0.00,0.00}{we can consider that most of} inelastic \textcolor{black}{contributions} come from the \textcolor{black}{projectile excitation}.

In \textcolor{black}{Fig.~\ref{fig:3}}, \textcolor{black}{cross sections are} determined without free parameters after fixing conditions, \textcolor{black}{such as excitation state energies and dipole strengths}.
Already, we have calculated inelastic cross section for $^{11}$Be+$^{197}$Au system \textcolor{black}{in the} same manner, which has shown good agreement with experimental data in Ref.~\cite{PhysRevC.92.044618}.
\textcolor{black}{Both results} at $E_{\textrm{c.m}} = 29.6$ MeV and $E_{\textrm{c.m}} = 37.1$ MeV also show reasonable agreement with experimental data without free parameters.
These results \textcolor{black}{are employed for} the results of QE scattering presented in Fig.~\ref{fig:2} by \textcolor{black}{adding} the elastic scattering results in Fig.~\ref{fig:1}.

\textcolor{black}{Here,} we \textcolor{black}{cannot} distinguish any \textcolor{black}{differences} the results between \textcolor{black}{the results} with \textcolor{black}{and} without LRN potential. Although partial wave function, $\chi^{(+)}_{l} (r)$, \textcolor{black}{is influenced by the real part in} LRN potential, real parts for LRN potential \textcolor{black}{are very small compared to \textcolor[rgb]{0.00,0.00,0.00}{those for the bare potential} $V_{\textrm{sh}}$} as presented in Table.~\ref{parameters_11be_r}. Therefore, in Fig.~\ref{fig:3}, all inelastic scattering results show \textcolor{black}{almost} same \textcolor{black}{results} independently of the LRN potential \textcolor{black}{type}.


\subsection{Breakup cross sections}
\label{sec:org976d646}
\begin{figure}[h]
\begin{tabular}{cc}
\includegraphics[width=0.50\linewidth]{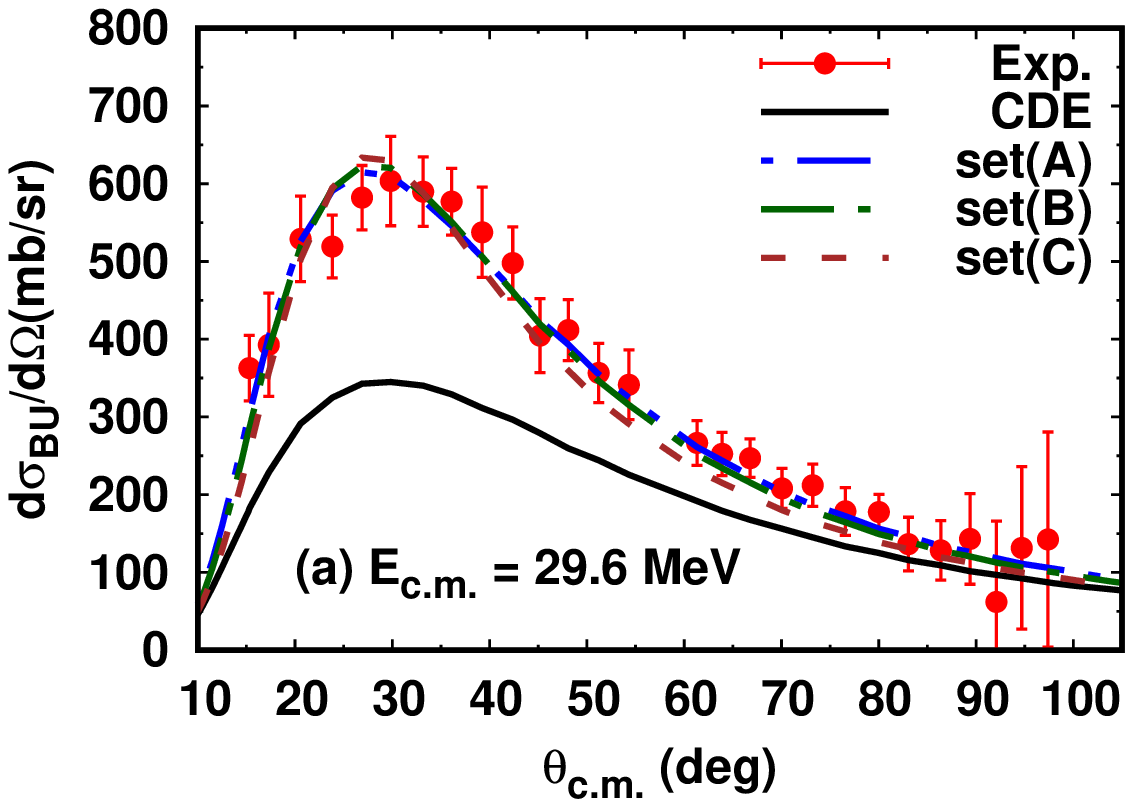} & \includegraphics[width=0.50\linewidth]{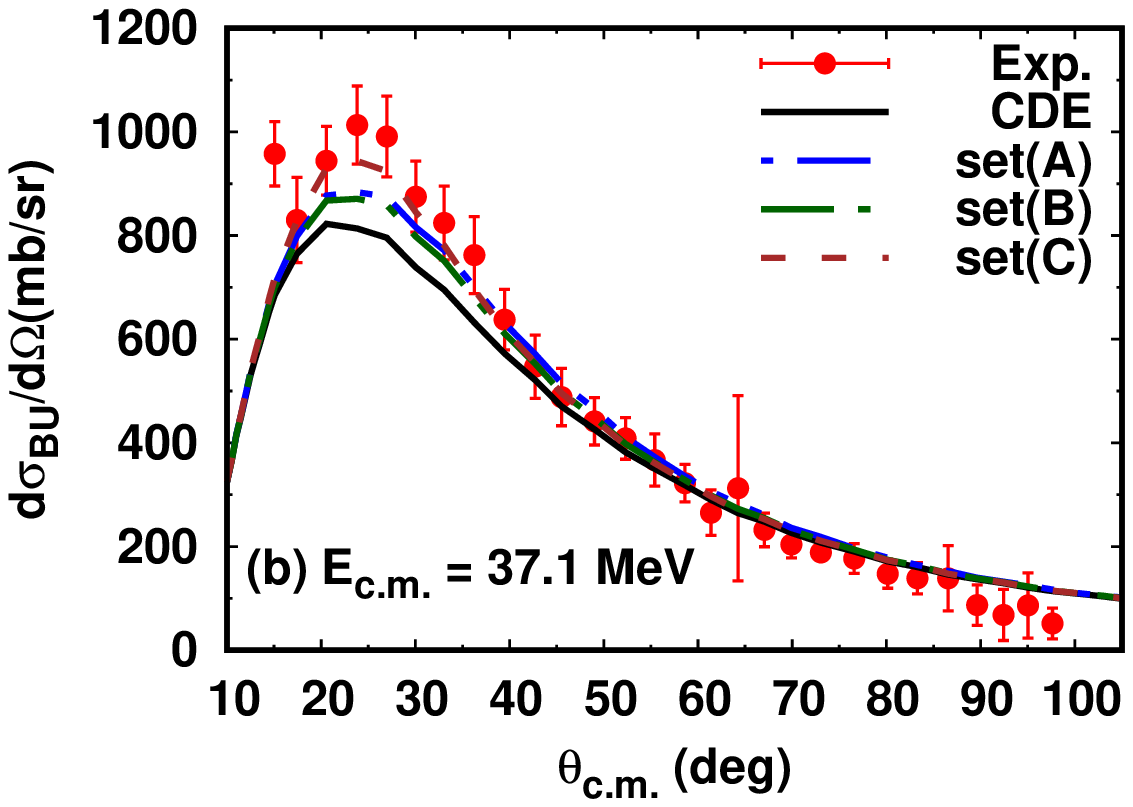} \tabularnewline
\end{tabular}
\caption{ (Color online)
The measured breakup cross section of $^{11}$Be + $^{197}$Au system at $E_{\textrm{c.m.}}$ = 29.6 MeV(\textcolor{black}{a}) and $E_{\textrm{c.m.}}$ = 37.1 MeV (\textcolor{black}{b}). The experimental \textcolor{black}{data are taken from} Ref.~\cite{PhysRevLett.118.152502}. The CDE means cross sections only with the CDE potential.}
\label{fig:4}
\end{figure}

In Fig.~\ref{fig:4}, we \textcolor{black}{present} results of breakup cross sections as \textcolor{black}{a} function of c.m. angle using \textcolor{black}{the} CDE potential in Eq.~(\ref{potential_coulomb-break}) \textcolor{black}{with and without   LRN potential.}

\textcolor{black}{In Fig.~\ref{fig:4},} \textcolor{black}{the effect of LRN potential, which is not seen in elastic scattering and quasi-elastic scattering cross sections, clearly appeared in the breakup reaction one.}
In \textcolor{black}{Fig.~\ref{fig:4}(a)}, the breakup cross sections are calculated at \textcolor{black}{$E_{\textrm{c.m}}$ =} 29.6 MeV \textcolor{black}{which} is slightly below Coulomb barrier.
\textcolor{black}{One can easily notice} that the results \textcolor{black}{including only CDE potential} without LRN potential \textcolor{black}{in} (black) solid line \textcolor{black}{definitely underestimated} experimental data.
It means that CDE potential for breakup reaction including B(E1) distribution is not sufficient to describe experimental \textcolor{black}{data} in this energy region. \textcolor[rgb]{0.00,0.00,0.00}{As a result, we need additional LRN potential in Eq.(\ref{surface-woods})}.
\textcolor{black}{Note that resutls by the all parameter sets of LRN potential in Table~\ref{parameters_11be_r} obtained from the simultaneous
$\chi^2$ analysis show} good agreement with experimental data.
The differences between the sets, (A), (B) and (C) are almost negligible \textcolor{black}{for the case at} \textcolor{black}{$E_{\textrm{c.m}}$ =} 29.6 MeV.

In \textcolor{black}{Fig.~\ref{fig:4}(b)}, we \textcolor{black}{show the results} at \textcolor{black}{$E_{\textrm{c.m}}$ =} 37.1 MeV\textcolor{black}{, which} are around Coulomb barrier.
\textcolor{black}{Cross sections} \textcolor{black}{including only} CDE effect without LRN potential \textcolor{black}{in} (black) solid line \textcolor{black}{are} larger than the case of \textcolor{black}{$E_{\textrm{c.m}}$ =} 29.6 MeV.
However, \textcolor{black}{they} still underestimated \textcolor{black}{experimental} data.
It mean that \textcolor{black}{one needs the} additional interaction like LRN potential.
In this case, we could find the difference \textcolor{black}{of the breakup cross section calculated by} parameter sets, (A), (B) and (C).
The set(C) \textcolor{black}{in} (brown) dashed line \textcolor{black}{is closest to the experimental data}.
We note that set(C) is the parameter set with \textcolor{black}{longest} radius $r^{\textrm{lo}}_{0}$ = $r^{\textrm{lo}}_{W}$ = 3.5fm in our \textcolor{black}{scheme}.
\textcolor{black}{It implies} that \textcolor{black}{we} have to consider long-range interaction potential in order to describe experimental data \textcolor{black}{relevant to the system including} halo nuclei.
\textcolor{black}{However, one can not definitely determine} the absolute radius of halo nuclei because this radius is \textcolor{black}{a} relative scattering length.

Results by the two sets, (A) and (B), denoted as dot-dashed (blue) and dashed-dot (green) line, have almost equal cross sections, and \textcolor{black}{underestimated} experimental data although elastic and inelastic cross \textcolor{black}{section data} are \textcolor{black}{well explained}.
Also, in \textcolor{black}{Table}~\ref{parameters_11be_r}, we notice that diffuseness parameter $a^{\textrm{lo}}_{0} = a^{\textrm{lo}}_{W}$ \textcolor{black}{decreases} by increasing radius because they \textcolor{black}{are intertwined}. But \textcolor{black}{they are} still huge as 6.24 at \textcolor{black}{$E_{\textrm{c.m}}$ =} 29.6 MeV.

\subsection{Energy dependency of CDE potential}
\label{sec:org519c882}
\textcolor{black}{In Fig.~\ref{fig:4}}, we have \textcolor{black}{presented} the contribution of CDE potential \textcolor{black}{for each} incident energy, \textcolor{black}{as} shown with (black) solid lines of both \textcolor{black}{at (a) $E_{\textrm{c.m}}$ = 29.6 MeV} and \textcolor{black}{(b) $E_{\textrm{c.m}}$ = 37.1 MeV.} 
In this calculation, we also calculated the breakup cross section only with CDE potential by switching off the LRN potential.
\textcolor{black}{Note that we could see that the contribution by the CDE potential increases with the increase of the incident energy, $E_{\textrm{c.m}}$ if we compare the breakup cross sections at} \textcolor{black}{$E_{\textrm{c.m}}$ =} 29.6 MeV and \textcolor{black}{$E_{\textrm{c.m}}$ =} 37.1 MeV.

\begin{figure}[h]
	\begin{center}
		\includegraphics[width=0.48\linewidth]{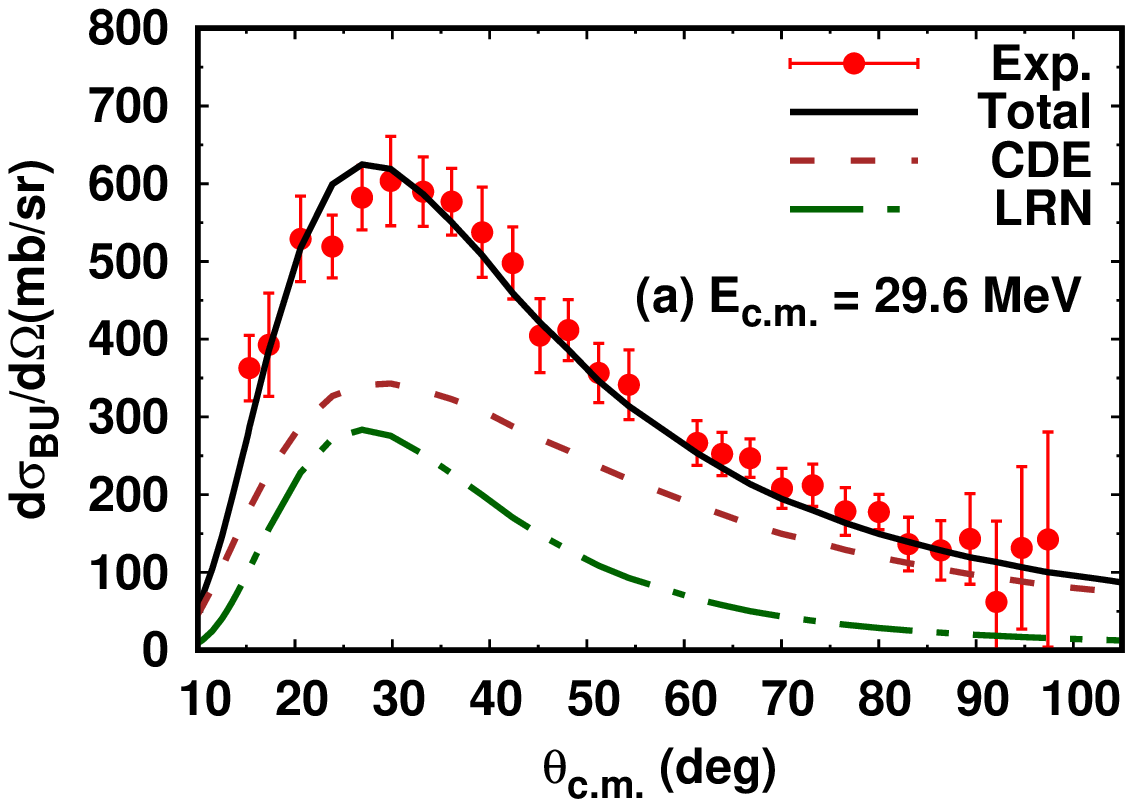}
		\includegraphics[width=0.48\linewidth]{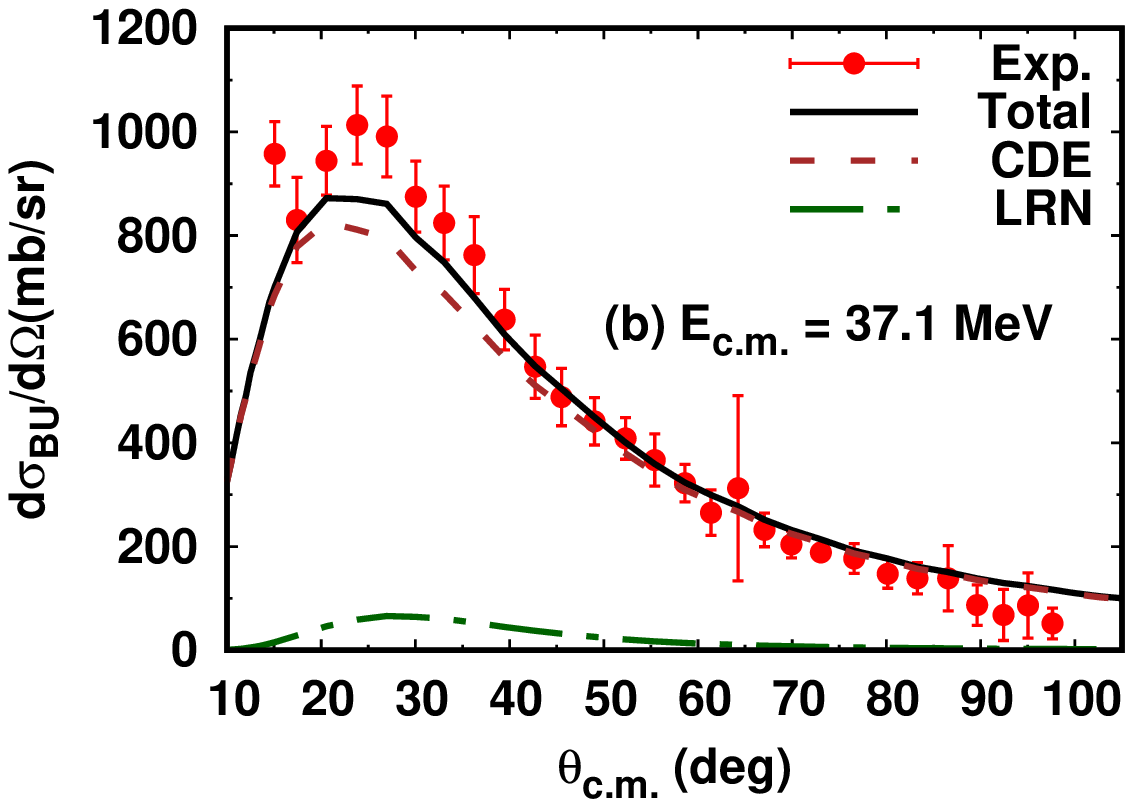}
	\end{center}
	\caption{(Color online)
		Angular distributions of breakup cross section $d\sigma_{\textrm{BU}}/d\Omega$ \textcolor{black}{at (a) $E_{\textrm{c.m}}$ = 29.6 MeV and (b) $E_{\textrm{c.m}}$ = 37.1 MeV, respectively.}
		Red and blue circles represent the experimental \textcolor{black}{data} \textcolor{black}{for $^{11}$Be + $^{197}$Au system~\cite{PhysRevLett.118.152502}. The (black) solid, (brown) dashed, and (green) dash-dotted lines show the angular distributions of breakup cross sections with the CDE + LRN (= Total) potential, only with the CDE, and only with LRN, reapectively.}}
	\label{fig:5}
\end{figure}

\textcolor{black}{To quantitatively understand the increased CDE potebtial effect along with the increased incident energy, however, we need to analyze the LRN and CDE effects on incident energy in the same frame {\it i.e.} with the LRN potential switched on. In Fig.~\ref{fig:5},
one can notice that experimental and theoretical results for breakup cross section are increased at forward angle  by increasing the incident energy.}
Note that main peaks of breakup cross section at forward angle are caused by the long-range Coulomb interaction {\it i.e.} CDE potential.

Particularly it is noteworthy that the contribution of the CDE potential increases significantly with the increase of the incident energy. This implies that most of the breakup reaction cross section at the incident energy around and above the Coulomb barrier are due to the effect of the CDE potential.
\subsection{Necessity of LRN potential}
Finally, we \textcolor{black}{argue regarding} the necessity of LRN potential \textcolor{black}{depending on the} incident energy.
\textcolor{black}{As shown in Fig.~\ref{fig:5}, the CDE potential} \textcolor{black}{itself for breakup} is not sufficient to describe experimental data \textcolor{black}{at the incident energy} below the Coulomb barrier.
It means that  we have to consider \textcolor{black}{an} additional potential like LRN \textcolor{black}{potential} \textcolor{black}{at the incident energy below} Coulomb barrier. This additional potential is thought to come from complicated interactions.
As mentioned in previous sections, the LRN potential \textcolor{black}{which takes implicitly into account} the interaction between a valance neutron part of projectile and target \textcolor{black}{nucleus}, is caused from halo structure of projectile.
\textcolor{black}{Unfortunately,} by the limitation of the present optical model, we could not explicitly describe \textcolor{black}{the interaction between a valance neutron part of projectile and target nucleus. However, one can easily notice that} they are implicitly taken into account by the contribution of breakup reaction \textcolor{black}{through the imaginary part of the LRN potential} in the optical model approach.

Additionally, the present theoretical results  in Fig.~\ref{fig:5} shows \textcolor{black}{a} sharply increasing tendency of the CDE contribution with the increase of the incident energy.
If incident energy is sufficiently large, we could conjecture that CDE potential for breakup would cover the LRN potential effect.
Then, there would be no \textcolor{black}{rooms for the} LRN potential anymore \textcolor{black}{above} certain critical energy point.
It means \textcolor{black}{that} specific halo properties \textcolor{black}{derived from the effects of the weakly bounded valence neutron(s)} would not work anymore, because the LRN potential contributes to describing properties of halo projectile in our description.
\textcolor{black}{Therefore,} the results in Fig.~\ref{fig:5} definitely shows that LRN potential we employed in this calculation plays vital roles of describing scattering experimental data relevant to halo \textcolor{black}{nuclei} \textcolor{black}{at the incident energy below Coulomb barrier.}

\section{Summary and Conclusions}
\label{sec:org4740eac}
We have simultaneously calculated elastic, quasi-elastic and breakup cross sections of $^{11}$Be + $^{197}$Au system\textcolor{black}{, where} $^{11}$Be has a halo structure\textcolor{black}{,} by taking into account long range dynamic polarization potentials, such as Coulomb dipole excitation (CDE) and long range nuclear (LRN) potentials, to the short range nuclear (SRN) potential in \textcolor{black}{an} extended optical model (OM) approach. In the present OM scheme, four free parameters in $\chi^{2}$ analysis have been used for understanding the whole data: elastic, inelastic and breakup cross \textcolor{black}{section data}, at 29.6 and 37.1 MeV.

First, we have successfully reproduced recent experimental data for elastic and quais-elastic scattering around Coulomb barrier, which data were given in terms of the ratio of the elastic and quasi-elastic scattering to Rutherford scattering data. CDE and SRN potentials turned out to be main potentials for properly accounting for most of the elastic and quasi-elastic scattering data. Although  the LRN potential contribution was found to be minor, there still remained some rooms for the necessity of LRN potential in the forward scattering region.

Second, therefore, we have \textcolor{black}{studied carefully} the energy dependence in \textcolor{black}{the angular distribution data for} breakup reaction \textcolor{black}{in order to} check the influence of LRN as well as CDE potential. In the low energy region around Coulomb barrier, CDE potential turned out to be insufficient for describing breakup cross sections, specifically, in the forward angle region.
This clearly indicates that the scattering data relevant to halo nuclei around Coulomb barrier needs \textcolor{black}{another type's} long range potential, {\it i.e.} LRN potential, stemming from the residual interactions peculair to halo nuclei scattering \textcolor{black}{in the present OM analysis}. But the LRN contribution becomes the smaller with the larger incident energy above Coulomb barrier.

\textcolor{black}{In conclusion, the long range dynamic polarization potentials, such as CDE and LRN potential, are shown to be necessary for properly describing low energy ion scattering by halo nuclei. But, for further definite conclusion for the LRN potential,} one needs more experimental data with halo nuclei projectiles especially around Coulomb barrier \textcolor{black}{to be expected from many radioisotope accelarator facilities in near future}.

\section*{Acknowledgment}
\label{sec:orgae7f2b2}
This work was supported by the National Research Foundation of Korea (Grant Nos. NRF-2016R1C1B1012874, NRF-2017R1E1A1A01074023, and 2018R1D1A1B07045915).


\begin{thebibliography}{}
\bibitem{satchler1983direct}
G. R. Satchler, Direct nuclear reactions, Oxford Univ. press (1983).
\bibitem{love1977dynamic}
W. Love, T. Terasawa and G. Satchler, Nuclear Physics A {\bf 291}, 183 (1977).
\bibitem{PhysRevC.75.024610}
W. Y. So, T. Udagawa, K. S. Kim, S. W. Hong and B. T. Kim, Phys. Rev. C {\bf 75}, 024610 (2007).
\bibitem{PhysRevC.77.024609}
W. Y. So, T. Udagawa, S. W. Hong and B. T. Kim, Phys. Rev. C {\bf 77}, 024609 (2008).
\bibitem{PhysRevC.90.054615}
W. Y. So, K. S. Kim, K. S. Choi and M.-K. Cheoun, Phys. Rev. C {\bf 90}, 054615 (2014).
\bibitem{PhysRevC.92.014627}
W. Y. So, K. S. Kim, K. S. Choi and M.-K. Cheoun, Phys. Rev. C {\bf 92}, 014627 (2015).
\bibitem{PhysRevC.89.057601}
W. Y. So, K. S. Kim and M.-K. Cheoun, Phys. Rev. C {\bf 89}, 057601 (2014).
\bibitem{PhysRevC.92.044618}
W. Y. So, K. S. Choi, M.-K. Cheoun and K. S. Kim, Phys. Rev. C {\bf 92}, 044618 (2015).
\bibitem{PhysRevC.93.054624}
W. Y. So, K. S. Choi, M.-K. Cheoun and K. S. Kim, Phys. Rev. C {\bf 93}, 054624 (2016).
\bibitem{AndresGomez-CamachoNagarajan1994}
M. Andr{\'e}s, J. G{\'o}mez-Camacho and M. Nagarajan, Nuclear Physics A {\bf 579}, 273 (1994).
\bibitem{Gomez-CamachoAndresNagarajan1994}
J. Gomez-Camacho, M. Andres and M. Nagarajan, Nuclear Physics A {\bf 580}, 156 (1994).
\bibitem{Andres199782}
M. Andr\'es, J. Christley, J. G\'omez-Camacho and M. Nagarajan, Nuclear Physics A {\bf 612}, 82 (1997).
\bibitem{TANIHATA1988592}
I. Tanihata, T. Kobayashi, O. Yamakawa, S. Shimoura, K. Ekuni, K. Sugimoto, N. Takahashi, T. Shimoda and H. Sato, Physics Letters B {\bf 206}, 592 (1988).
\bibitem{PhysRevC.63.061603}
E. F. Aguilera, J. J. Kolata, F. D. Becchetti, P. A. DeYoung, J. D. Hinnefeld, \'A. Horv\'ath, L. O. Lamm, H.-Y. Lee, D. Lizcano, E. Martinez-Quiroz, P. Mohr, T. W. O'Donnell, D. A. Roberts and G. Rogachev, Phys. Rev. C {\bf 63}, 061603 (2001).
\bibitem{PhysRevC.82.044617}
A. Lemasson, A. Navin, N. Keeley, M. Rejmund, S. Bhattacharyya, A. Shrivastava, D. Bazin, D. Beaumel, Y. Blumenfeld, A. Chatterjee, D. Gupta, G. de France, B. Jacquot, M. Labiche, R. Lemmon, V. Nanal, J. Nyberg, R. G. Pillay, R. Raabe, K. Ramachandran, J. A. Scarpaci, C. Simenel, I. Stefan and C. N. Timis, Phys. Rev. C {\bf 82}, 044617 (2010).
\bibitem{PhysRevLett.110.142701}
J. P. Fern\'andez-Garc\'{\i}a, M. Cubero, M. Rodr\'{\i}guez-Gallardo, L. Acosta, M. Alcorta, M. A. G. Alvarez, M. J. G. Borge, L. Buchmann, C. A. Diget, H. A. Falou, B. R. Fulton, H. O. U. Fynbo, D. Galaviz, J. G\'omez-Camacho, R. Kanungo, J. A. Lay, M. Madurga, I. Martel, A. M. Moro, I. Mukha, T. Nilsson, A. M. S\'anchez-Ben\'{\i}tez, A. Shotter, O. Tengblad and P. Walden, Phys. Rev. Lett. {\bf 110}, 142701 (2013).
\bibitem{Cubero2012}
M. Cubero, J. P. Fern\'andez-Garc\'{\i}a, M. Rodr\'{\i}guez-Gallardo, L. Acosta, M. Alcorta, M. A. G. Alvarez, M. J. G. Borge, L. Buchmann, C. A. Diget, H. A. Falou, B. R. Fulton, H. O. U. Fynbo, D. Galaviz, J. G\'omez-Camacho, R. Kanungo, J. A. Lay, M. Madurga, I. Martel, A. M. Moro, I. Mukha, T. Nilsson, A. M. S\'anchez-Ben\'{\i}tez, A. Shotter, O. Tengblad and P. Walden, Phys. Rev. Lett. {\bf 109}, 262701 (2012).
\bibitem{PhysRevC.87.044603}
A. M. Vinodkumar, W. Loveland, R. Yanez, M. Leonard, L. Yao, P. Bricault, M. Dombsky, P. Kunz, J. Lassen, A. C. Morton, D. Ottewell, D. Preddy and M. Trinczek, Phys. Rev. C {\bf 87}, 044603 (2013).
\bibitem{PhysRevC.92.044608}
J. P. Fern\'andez-Garc\'{\i}a, M. Cubero, L. Acosta, M. Alcorta, M. A. G. Alvarez, M. J. G. Borge, L. Buchmann, C. A. Diget, H. A. Falou, B. Fulton, H. O. U. Fynbo, D. Galaviz, J. G\'omez-Camacho, R. Kanungo, J. A. Lay, M. Madurga, I. Martel, A. M. Moro, I. Mukha, T. Nilsson, M. Rodr\'{\i}guez-Gallardo, A. M. S\'anchez-Ben\'{\i}tez, A. Shotter, O. Tengblad and P. Walden, Phys. Rev. C {\bf 92}, 044608 (2015).
\bibitem{PhysRevLett.105.022701}
A. Di Pietro, G. Randisi, V. Scuderi, L. Acosta, F. Amorini, M. J. G. Borge, P. Figuera, M. Fisichella, L. M. Fraile, J. Gomez-Camacho, H. Jeppesen, M. Lattuada, I. Martel, M. Milin, A. Musumarra, M. Papa, M. G. Pellegriti, F. Perez-Bernal, R. Raabe, F. Rizzo, D. Santonocito, G. Scalia, O. Tengblad, D. Torresi, A. M. Vidal, D. Voulot, F. Wenander and M. Zadro, Phys. Rev. Lett. {\bf 105}, 022701 (2010).
\bibitem{PhysRevC.85.054607}
A. Di Pietro, V. Scuderi, A. M. Moro, L. Acosta, F. Amorini, M. J. G. Borge, P. Figuera, M. Fisichella, L. M. Fraile, J. Gomez-Camacho, H. Jeppesen, M. Lattuada, I. Martel, M. Milin, A. Musumarra, M. Papa, M. G. Pellegriti, F. Perez-Bernal, R. Raabe, G. Randisi, F. Rizzo, G. Scalia, O. Tengblad, D. Torresi, A. M. Vidal, D. Voulot, F. Wenander and M. Zadro, Phys. Rev. C {\bf 85}, 054607 (2012).
\bibitem{PhysRevLett.118.152502}
V. Pesudo, M. J. G. Borge, A. M. Moro, J. A. Lay, E. N\'acher, J. G\'omez-Camacho, O. Tengblad, L. Acosta, M. Alcorta, M. A. G. Alvarez, C. Andreoiu, P. C. Bender, R. Braid, M. Cubero, A. Di Pietro, J. P. Fern\'andez-Garc\'{\i}a, P. Figuera, M. Fisichella, B. R. Fulton, A. B. Garnsworthy, G. Hackman, U. Hager, O. S. Kirsebom, K. Kuhn, M. Lattuada, G. Marqu\'{\i}nez-Dur\'an, I. Martel, D. Miller, M. Moukaddam, P. D. O'Malley, A. Perea, M. M. Rajabali, A. M. S\'anchez-Ben\'{\i}tez, F. Sarazin, V. Scuderi, C. E. Svensson, C. Unsworth and Z. M. Wang, Phys. Rev. Lett. {\bf 118}, 152502 (2017).
\bibitem{PhysRevC.65.044616}
B. T. Kim, W. Y. So, S. W. Hong and T. Udagawa, Phys. Rev. C {\bf 65}, 044616 (2002).
\bibitem{Kakuee2003339}
O. Kakuee, J. Rahighi, A. S\'anchez-Ben\'itez, M. Andr\'es, S. Cherubini, T. Davinson, W. Galster, J. G\'omez-Camacho, A. Laird, M. Lamehi-Rachti, I. Martel, A. Shotter, W. Smith, J. Vervier and P. Woods, Nuclear Physics A {\bf 728}, 339 (2003).
\bibitem{MAHAUX1986354}
C. Mahaux, H. Ngô and G. Satchler, Nuclear Physics A {\bf 449}, 354 (1986).
\bibitem{PhysRevC.69.064606}
W. Y. So, S. W. Hong, B. T. Kim and T. Udagawa, Phys. Rev. C {\bf 69}, 064606 (2004).
\bibitem{PhysRevC.69.047601}
J. J. Kolata, E. F. Aguilera, F. D. Becchetti, Y. Chen, P. A. DeYoung, H. Garc\'{\i}a-Mart\'{\i}nez, J. D. Hinnefeld, J. H. Lupton, E. Martinez-Quiroz and G. Peaslee, Phys. Rev. C {\bf 69}, 047601 (2004).
\bibitem{NAKAMURA1994296}
T. Nakamura, S. Shimoura, T. Kobayashi, T. Teranishi, K. Abe, N. Aoi, Y. Doki, M. Fujimaki, N. Inabe, N. Iwasa, K. Katori, T. Kubo, H. Okuno, T. Suzuki, I. Tanihata, Y. Watanabe, A. Yoshida and M. Ishihara, Physics Letters B {\bf 331}, 296 (1994).
\bibitem{borge2017scattering}
G. Borge, J. Maria, V. Pesudo, E. N{\'a}cher, A. Perea, A. Moro, J. A. Lay, M. Alvarez, J. P. Fernandez-Garcia, A. Di Prieto and others, The 26th International Nuclear Physics Conference (proceeding of science)  {\bf 281}, 207 (2017).
\textcolor[rgb]{0.00,0.00,0.00}{\bibitem{ajze98}
F. Ajzenberg-Selove, Nucl. Phys. A {\bf 506}, 1 (1998).}
\end{thebibliography}
\end{document}